\documentclass[12pt]{article}
\usepackage{latexsym}
\usepackage[english]{babel}
\usepackage{fancyhdr}
\usepackage[mathscr]{eucal}
\usepackage[dvips]{graphicx}
\usepackage{graphics}
\usepackage{color}
\usepackage{epsfig}
\usepackage{amsmath}
\usepackage{amsthm}
\usepackage{amsfonts}
\usepackage{amssymb}
\usepackage{amscd}
\usepackage{bbm}
\usepackage{latexsym}
\usepackage{marvosym} 
\usepackage{pifont}
\usepackage{subfigure}
\usepackage{psfrag}
\usepackage{caption}

\begin{document}

\title{Zero range (contact) interactions conspire to produce  Efimov trimers and quadrimers}

\maketitle
G.F. Dell'Antonio

Math. Dept. University Sapienza (Roma)

and

Mathematics Area, Sissa (Trieste)

\vskip 1 cm \noindent

Summary

We introduce \emph{contact  (zero range) interactions } , a special class of self-adjoint extensions of the  N-body Schr\"odinger free hamiltonian $ H_0$ restricted to functions with support away from the \emph{contact  manyfold} $ \Gamma \equiv \cup \Gamma_{i,j} \;\;\; \Gamma_{i,j}\equiv \{x_i = x_j \;  i \not= j \} \;,\; x_i \in R^3 $.

These  extensions are defined by boundary conditions at $ \Gamma$.

We discuss the spectral properties as function of the masses and the statistics

The  (Efimov) spectrum is entirely due "conspiracy" of the contact interactions of two  pairs. 
 
These states are called in  Theoretical Physics  \emph{trimers} if the two  pairs have an element in common, \emph{quadrimers} otherwise. 

The analysis can be extended to the case in which there is a regular two-body potential, but then the spectral properties cannot be given explicitly.
 
We  prove that these interactions are limit (in strong resolvent sense) when  $ \epsilon \to 0$ of  N-body hamiltonians $ H_\epsilon = H_0+ \sum_{i,j} \frac {1}{\epsilon^3}V_{i,j}( \frac {|x_i-x_j|}{\epsilon})  ,\; V_{i,j} \in L^1 $.

The result is \emph{ independent of the shape of the potential}

Strong resolvent convergence implies convergence of spectra.This makes contact interaction a valuable tool in the study of the spectrum of a system of $N$ particles  interacting through potentials of very short range.

\section {Introduction} 

We introduce \emph{contact interactions} as a special type of self-adjoint extensions of the operator $ \hat H_0 $ defined  as the free Schr\"od"nger operator $H_0$  with masses $m_i$ restricted  to functions that have support away from  $ \Gamma \equiv  \cup' \Gamma_{i,j} \;\; \Gamma_{i,j} \equiv \{x_i = x_j \}\;  i \not= j \; x_i \in R^3 $ .

The prime indicates that only some pairs of indices are present. 

The self-adjoint extensions we consider are characterized by boundary conditions  at $ \Gamma$. 

At $ \Gamma_{i,j}$ we admit a singularity $\frac {C_{i,j} }{|x_i - x_j|} $ for functions in the domain  (some of the $ C_{i,j} $ can be zero) .

Imposing this behavior at $ \Gamma$ is  formally  equivalent to placing a (negative) distributional potential at the boundary; this is seen by taking the scalar product with functions in the range  of $\hat H_0 $ and integrating by parts twice  (the boundary terms vanish) .

Notice that functions in the domain of $H_0$  are absolutely continuous; therefore  they are in the form domain of $H_0$ \emph{and of the boundary potential}.

Conditions  of this type were imposed  by H.Bethe and R.Peirels [B,P] in 1935 in their analysis of the proton-neutron interaction; later they were used  for the case $N=3 $ by G. Skorniakov and K.Ter Martirosian  [S,T] in their study of the three body problem in Nuclear Physics in the  Faddaev formalism . 

For every $N$ we shall denote them  Ter Martirosian -Skorniakov (T-S) boundary conditions .

The analysis we make can be extended to any Schr\"odinger hamiltonian bounded below with regular potentials. Here we consider only the case of the free hamiltonian $H_0$

We  prove that hamiltonians with contact interactions are limits in strong resolvent sense when  $ \epsilon \to 0$ of hamiltonians $ H_\epsilon= H_0 +  \sum_{i,j} V^\epsilon_{i,j}   (|x_j -x_i)  $ with   two-body negative potentials  which scale according to $ V^\epsilon (|x|) =\epsilon ^{-3} V ( \frac {|x|}{ \epsilon} )$ where $ V(|x|) \in L^1 $.

 Notice that  the $L^1 $ norm of $ V^\epsilon$  does not depend on $ \epsilon$ and that the result is independent of the shape of the potential.

This makes contact interactions a valuable tool in the the study of the spectrum of a system of $N$ particles of mass $m_i$ interacting through potentials of very short range.

\bigskip
\emph{Remark 1}

Conspiracy of contact interaction is the counterpart  of conspiracy of zero energy resonances [K,S]. 

Resonances may  be viewed as \emph{contacts in a point at infinity} (direction of slow decrease in the difference of the coordinates).

Since the Laplace operator in dimension three is conformally covariant,   a pair of particles contact interaction  with a zero energy resonance can be considered to be equivalent to contact interaction between two pairs of particles (one pair at infinity).   

Recall that zero energy resonances between two particles correspond to a  $ \frac {1}{ |x_- x_j|}$  behavior of the wave function at infinity;  this is the same behavior (at $ \Gamma_{i,j} $)  that we have required in case of  contact interactions.

Conspiracy may occur  between resonances  [K,S]; this  leads to the production of an Efimov set of  bound states. 

The analogous  bound  states  for contact interactions are called \emph{trimers} and \emph{quadrimers} in the Physics literature . 
 
 \bigskip
 ...............................  

\bigskip
\emph{Remark 2}

A \emph{ different }  hamiltonian of "zero range  interaction"  was introduced [A] and called  \emph{point interaction}.

The system is composed of one particle in interaction with a fixed point trough a potential that has a zero-energy resonance. 

In an appendix we comment on the  relation  between contact and point interactions . 

In particular we prove that point interactions  are limit of a contact  interaction for a system of three identical bosons (two at infinity) of mass one and a particle of mass $M$,  when $ M \to \infty $ (so that this particle in the limit is a fixed point) and the scaling factor of the potential is that given in [A] i.e $ V^\epsilon  (|y|) = \frac {1}{ \epsilon^2} V(\frac {|y|}{\epsilon})$.
 
If the spectrum is positive, the spectral measure has a singularity at zero.

If the spectrum is not positive we  show that the entire negative part of the spectrum of the point interaction hamiltonian \emph{ is singular continuous}.

It is the limit of a four-body Efimov point spectrum, when the distance between adjacent  points of the spectrum goes to zero.

In the limit each point of the negative part of the spectrum is an accumulation point of the spectrum of the "point interaction" hamiltonian for $ M < \infty $.

This "explains" the mapping properties of the Wave operator for point interactions [D].

We show also that a "simpler explanation" is obtained by a change of time scale. 

In the appendix  we  also mention briefly the relation of our approach to the "heat kernel renormalization"  for the three-body problem [E,T]

\section {Auxiliary space}

Contact interactions are a special class of self-adjiont extensions of the symmetric operator $ \hat H_0$ defined to be the free hamiltonian restricted to functions that vanish in a neighborhood of $ \tilde \Gamma$.

It will convenient to introduce an auxiliary space, suggested (as we will see)   by the special class of self-adjoint extensions we consider. 

For the three-body case our approach follows the lines of the approach of R.Minlos ($ [M_1][M_2] $ , see also $[C_1][C_2]$ ).

For this reason we call the auxiliary space \emph{Minlos space} and denote it by the symbol $ { \cal M}$.

The "physical space" $L^2(R^{3N}) $   is compactly embedded in $ {\cal M} $  by a map that we will give explicitly. 

We call this map \emph{Krein map} and we denote it by the symbol $ {\cal K}$.

The extensions of positive symmetric operators have been studied in great detail by Birman, Visik and Krein [B][K]. We use a quadratic form  version described in [A,S].

In  $ {\cal M} $  the boundary conditions at $ \Gamma $ are represented by $ L^1$ potentials  and in this space the asymptotic convergence  as $ \epsilon \to 0$  is easily proved.

In $ { \cal M}$ we study in some detail the case of three particles, in particular the interaction of two identical particles of mass one with a third one, paying attention to the dependence on the mass of the third particle and to the statistics.

We consider also in some detail the four-particle case, in particular  the case of two  pairs of identical particles, either fermions or bosons.

Finally we consider the case of $N$ pairs of identical particles, both in the case of fermions and in the case of bosons. 

In $ { \cal M }$  both the kinetic energy  the potential are essentially self-adjoint  unbounded operators; one is positive the other negative (we have assumed that the interaction is attractive). 

Their sum is a priori only symmetric. 

In the case we are considering the "boundary potential"  has a special structure in $ \cal{ M}$: it is the sum of a negative Coulomb potential and a positive regular term. 

The "kinetic term" is a positive (pseudo)-differential operator of order one (the square root of the free hamiltonian) 

For some choices of masses and symmetries the sum of the kinetic and potential terms defines in ${ \cal M} $ a unique (essentially) self-adjoint operator while for other choices there are continuous families of self-adjoint extensions (this  is an instance of Weyl limit circle property [D,R] ).

Depending on the masses and the symmetry properties,  the negative point spectrum of  each member of the family  may be absent  or contain a finite  or infinite set of  points. 

In the latter case the spectrum shows  the Thomas effect (geometric divergence of the eigenvalues to $-\infty$).

One has now to come back to the "physical space" $ L^2 (R^{3N}) .$ 

If there are no bound states this is done by inverting the Krein map.

It is then easy to prove \emph{in  physical space}  convergence when $ \epsilon \to 0$  \emph{in strong resolvent sense}  of the approximating hamiltonians (with shrinking potentials) to the hamiltonian with contact interactions. 

This follows because for positive quadratic forms weak convergence implies strong convergence and this in turn implies resolvent convergence. . 

If there are bound states in ${\cal M}$ showing the Thomas effect,  inverting the Krein map implies  a  difference in metric and the sequence of states is now an  \emph{Efimov sequence}: in physical space (the points in the negative  spectrum converge to zero geometrically).

One verifies that, after  reducing with respect to the continuous symmetries (our potentials are all rotationally invariant) , the quadratic form of the  potential in ${ \cal M} $ is \emph{strictly convex}.

A unique limit when $ \epsilon \to 0 $ is then obtained using convexity, compactness and lower semicontinuity (Gamma-convergence )[Dal].

Strong resolvent convergence follows from Gamma-convergence.

We prove that, depending on the masses and the symmetries, the conspiracy of two contact interactions  can give rise to three-body bound states (called \emph{trimers} in the Physics literature) and four-body bound states (\emph{quadrimers}).

The trimers' wave functions have essential support in a neighborhood of the triple coincidence point.

The eigenfunctions are obtained (by duality) applying the the Krein map to the eigenfunctions found in $ { \cal M}$ (the latter have a $ \frac {1}{|x_i - x_j|}$ singularity at the triple coincidence point).

The quadrimer's wave functions are centered in the quadruple coincidence point and are functions of the difference of the coordinates $X_i$ of  the barycenters of the two pairs. 

In $ { \cal M}$ they have a $ \frac {1}{|X_i - X_j|}$ singularity.They are obtained in physical space by applying the Krein map.

We consider briefly the N-body problem , $ N \geq 4 ,$ in particular in the case of pairs of identical particles (either fermions or bosons).

We show that in the  case of fermions the hamiltonian is positive (stability of the unitary Fermi  gas). 

In the case of identical bosons \emph{no new types of bound states occur} (the bound states are due to conspiracy of \emph{two} contact interactions ). 

The negative point spectrum is bounded by $-c N$  where the constant $c$ depends on the mass and on the strengthof the contact .

Presumably the  HKVZ theorem is valid for contact interactions as well as all threshold properties and Mourre estimates but we don't give an explicit  proof.

Since there are no singularities at the bottom of the continuous spectrum, the mapping properties of the wave operator are the usual ones [Y]

\bigskip
\emph{Remark 1 }

The strategy we use has much in common  with the method of \emph{boundary triplets} [ B,M,N][D,M]
which is used to find self-adjoint extensions of differential symmetric operators defined by excluding boundaries; the triplet is composed of the operator itself,  an operator at the boundary and the (generalized) Weyl function connecting the two. 

A basic example in electrostatics is given by the potential and the distribution of charges at the boundary.
The theory has been much generalized [D,H,M]

The Krein map (induced by a first order differential operator) has the  structure of the abstract Weyl function  [D,M].

We shall not elaborate here on this point.

\bigskip	
\emph{Remark 2} 

For $N=2$  \emph{the approach we use does not work}, the reason being that in the center of mass frame the contact manyfold reduces to a point and there is no room for a compact embedding. 

The "physical reason" is that  for $N=2$ there is  only one  contact interaction  and "conspiracy"  cannot  occur.

The same problem is encountered in electrostatics: the potential due to a "point charge" is well defined,
but in order to define a \emph{charge at a point}  one must consider the distribution of charges over a sphere of radius $ \epsilon$ and then let $ \epsilon \to 0$ ( dropping  a diverging term).

There is no problem in defining a charge distribution on a manyfold.

Notice  the limit equation is well defined also for $N=2$ (\emph{the limit equation does not depend on the metric of the space}).

In the case of two identical bosons the limit equation for a pair in contact interaction  is the Schr\"odinger equation with a cubic nonlinearity (interpreted as an equation for two identical particles in contact).

\section{Analytic formulation}

We study  in the limit $ \epsilon \to 0 $  a system of $N\geq 3$ particles in $R^3$  which satisfy  the 
Schr\"odinger equation with  hamiltonian 

\begin{equation}
H^\epsilon (V)  = -\sum_{k= 1}^N   \frac { 1} {2 m_k} \Delta_k + \sum_{i,j = 1 \ldots N } V_{i,j}^\epsilon (|x_1 - x_j|)
\end{equation}

where  the potentials scale as $ V_{i,j}^\epsilon (|y|) = \epsilon^{-3} V_{i,j} ( \frac {|y|}{\epsilon}) $  and $ V_{i,j} (x) $ are  (negative) $L^1$ functions. 

The sum is in general over a subset of the indices. 

Notice that the $L^1(R^3) $ norm of the potentials does not depend on $ \epsilon$ and the potentials  
  converges \emph{weakly} (in distributional sense) when $ \epsilon \to 0$ to a delta distribution at  the boundary $ \Gamma$.  

In the limit $ \epsilon \to 0$ the "potentials"  are therefore distributions supported by  \emph{some} of the contact hyper-planes $ \Gamma_{i,j} \equiv \{ x_i = x_j \}$ . 

We take the potential to be negative. 

 \emph{The limit  is described equivalently by  boundary conditions} at $\Gamma$ . 
 
 The equivalence can be seen by taking the scalar product with functions in the range of $\hat H_0$ and integrating by parts.
 
 This step is crucial if one wants to see contact interactions as limits of interactions with smaller and smaller support. 

 This is the same procedure which provides different realizations e.g. of the laplacian on $ (0, \infty) $
 
 We will investigate resolvent convergence and spectral properties. 
 
With our scaling in the  limit  $ \epsilon \to 0$ the potential is formally a distribution supported by some of the \emph{coincidence} hyper-planes $ \Gamma_{,i,j} \equiv \{x_i = x_j\} .$ 

The limit hamiltonian is therefore a self-adjoint extension of  $\hat  H_0$,   the free hamiltonian defined on functions that  have support away from the coincidence manyfold 

\begin{equation} 
\Gamma \equiv \cup'_{i,j} \Gamma_{i,j} \qquad \Gamma_{i,j} \equiv \{ x_i - x_j = 0 \} 
\end{equation} 

The prime over the sum indicates that  some of the hyper-planes do  not contribute to the interaction. 

 Functions in the domain of the extension may have a singularity $ \frac {1}{|x_i - x_j|} $ at $ \Gamma_{i,j}  $. 
 
\bigskip
We have remarked that contact interactions are extensions of the symmetric operator $\hat H_0$ defined as the  free hamiltonian  restricted to functions with support away from $ \Gamma$ [P]. 

Since this operator is positive, the possible extensions are classified by the theory of Birman, Krein, Visik [B][K].

To analyze \emph{the specific extension we consider} we use a quadratic-form version of the theory  [A,S] and introduce the auxiliary space ${ \cal M}$ in which the "physical space" is compactly embedded.

The  space $ { \cal M }$    is obtained  acting on $ L^2 (R^{3N} )$ with  $ (H_0 + \lambda)^{- \frac {1}{2}}$  where $H_0$ is the free hamiltonian. 

Here $ \lambda $ is arbitrary \emph{but strictly  positive}.  We  have called ${\cal K}$ this map.

Since to go back to physical space one inverts the map $ { \cal K} $, the actual value of the parameter $ \lambda > 0  $ is irrelevant.

The embedding has the advantage that the distributional "boundary potentials" are now less singular (they are $ L^1$ functions) and convergence when $ \epsilon \to 0$ is  easier to prove.

In $ {\cal M}$ the kernels of the quadratic forms are continuous in $ \lambda $ at $ \lambda = 0 $ and one can set $ \lambda = 0$ ; this simplifies considerably the analysis because the kernels are now  homogeneous of order $ -1$ in the coordinates. 

This helps greatly in keeping the formulae simpler and does not alter the small  distance behavior.  

\bigskip
We denote $H_M$ the operator in $ { \cal M }$   which is image of the hamiltonian under the map $ {\cal K}$.   

For the case $N=3$ the operator $H_M$   has been introduced   by R. Minlos $[M_1][M_2]$ (see also $[C_1$]) in his analysis of contact interactions.  

In an unpublished manuscript (private communication) R.Minlos attempted to treat in this way the case of two pairs of identical particles  but the formulation in momentum space (typical of the Fadeev formalism) prevented him to reach definite conclusions.

The operator $H_M$  is  the difference of two unbounded positive self-adjoint operators associated respectively to the kinetic energy and  to a  negative potential that has only Coulomb-type singularities.

Depending on the masses and the strength of the potential, the resulting operator may be essentially self-adjoint and positive ( "regular" case) or only symmetric and not positive ("singular" case).

In $[M_1][M_2]$  the multiplicity of self-adjoint is derived from the multiplicity of solutions of the equation $(\hat H_0^* +1)u = 0$. 

Here we work in configuration space and the multiplicity is seen as a Weyl limit circle property.

In the regular case  the proof of convergence in physical space is done by using the compactness of the map $ { \cal K}$ and  the fact that weakly closed  \emph{positive} quadratic forms are strongly closed.

In the singular case some of the components represent quadratic forms in a space of three particle, some in the space of four particles. 

Each component can be decomposed using invariance under rotations; after this decomposition  the quadratic form is \emph{strictly convex and lower semicontinuous}. 

The same properties hold for the quadratic forms of the approximating  hamiltoninans. 

This allows to use  Gamma- convergence [Dal]  when  $ \epsilon \to 0$; strong resolvent converge follows. 

This procedure selects \emph{uniquely} in each channel the limit operator: it is  the one with the lowest spectrum among the possible extensions. 

The resulting operator \emph{in physical space} has a negative point spectrum;  it may  consists of infinitely many points,  accumulating geometrically at zero (this difference between the spectral  properties in $ { \cal M }$  and in physical space is due to the difference in metric). 

This "Efimov effect"  is due  to conspiracy of contact interactions. Compare with the fact that  in the case of two body smooth interactions with a zero energy resonance this effect is due to conspiracy of zero-energy resonances  [K,S]).

This correspondence should  not come as a surprise since the two effects are both due to a $ \frac {1}{ |x_i- x_j|}$ behavior of functions at the boundary (i.e. at $ \Gamma$ for contact interactions, at the sphere at infinity in the case of resonances). 

The spectrum of the N-body system with contact interactions depends on the masses and on the statistics.  

We will prove that  for identical particels the (negative) point spectrum \emph{is completely determined} by the three- and four-particles sub-sectors ; this should be expected since the bound states are  due to "conspiracy of \emph{two} two-body contact interactions".

\bigskip
In order to exemplify the method, we analyze in detail the following cases: 

1) a pair of identical particles  of mass one interacting with a third particle of mass m; we consider  both the fermionic and the boson case.

2) two pairs of identical particles of mass 1, either fermions or bosons. 

3) N pairs of identical particles either fermions or bosons.

In the last case we prove that in the case of fermions (unitary gas) the hamiltonian is positive  for any value of $N$. 

In the case of bosons the (negative) lower bound of the spectrum is linear in $N$.

\bigskip
\emph{Remark 1} 

In the space $ { \cal M }$  the singularity at $\Gamma_{i,j} $ \emph{of the wave functions associated to the continuous part of the spectrum}  depends  on the position, masses and  symmetries  of all the particles. 

They are given now as 

\begin{equation}
\phi (X) = \sum_{i,j} \frac {a_{i,j} (Y ) }{ |x_i-x_j|} + \sum_{i,j}  b_{i,j} (Y) + o(\frac {1}{|x_i-x_j|} )
\end{equation}

where $Y$ represents the other differences of variables and $A \equiv \{ a_{i,j} \}, \;\  B \equiv  \{ b_{i,j} \}$ 
are suitable functions at the boundary.  

The functions $ a_{i,j} (Y) , b_{i,j}(Y)  $  depend on the masses of all the particles; this dependence is entirely due to the Krein map.

There are choices of the masses for which the matrix $ A \equiv a_{i,j}(Y) $ is singular (in some cases as singular as  $ \frac { 1}{|x_i -x_j | log(|x_i - x_j | )} $ in some  variable).  

Notice that these singular  boundary conditions occur in $ { \cal M}$;  \emph{in physical space one has the T-S boundary conditions on  function in the continuous spectrum}. 

\bigskip
{Remark 2} 

A stronger modification of the hamiltonian  occurs if one considers \emph{simultaneous} contact of three identical particles.  

Functions in the domain have now  \emph{in phyical space} a singularity $\frac  {1}{| x_i - x_j ||x_i - x_k| } $ at $ \Gamma$,  for  different values of the  indices.

 In [M,F] the authors consider this case for $N=3$ . It describes \emph{in physical space} three particles \emph{in simultaneous contact interaction}.

Also in this case  there is a family of dynamics (self-adjoint extensions). 

Each self-adjoint extension  is unbounded below \emph{in  physical space}  and  has infinitely many eigenvalues $ \mu_i ,  i \in Z.$  which diverge geometrically to $ -\infty$. 

\bigskip
\emph{Formally} these extensions could  be recovered as limit of hamiltonians  with smooth potential $V^\epsilon_{i,j}  (|x_i - x_j| ) $ which belong uniformly to $L^1$, have  support in a region of volume $ \epsilon ^3 $ and are  scaled so that the  $L^1$ norm stays constant.   

A rigorous proof is more difficult.

\bigskip

\section{Two identical scalar fermions and a different particle} 

For three particles  we will consider only the case in which a  particle of mass $m$  is in contact interaction with two identical particles of mass 1 while these two particles do not interact .

We consider first the case of two identical fermions.

The analysis in $ { \cal M }$  of the quadratic form of the operator  has reported in $ [M_1][M_2 ]$ [Pa] (see also $ [C_1][C_2]$)

Due to the symmetry one can consider a  quadratic form on a space of functions on $ R^3$.

Setting for simplicity $ \lambda = 0 $ in $ {\cal M}$ the form  is the sum of two terms 

\begin{equation}
 Q= Q_1+   Q_2
\end{equation} 
 
 \begin{equation}
 Q_1(\phi)  =  \frac {m}{m+1} (\phi, \sqrt {H_0}  \phi) 
\end{equation}

\begin{equation}
Q_2(p,q) = -\frac { \frac{2}{1+m} (p.q) } { (p^2 + q^2)^2 - \frac {4}{ (1 + m)^2 } (p.q)^2 } 
\end{equation}

These are the quadratic forms in $ { \cal M}$  that correspond respectively the kinetic energy and to the distributional potential.

These quadratic forms are invariant under rotations and therefore can be analyzed separately in each angular momentum sector. 

The mass $m$ of the third particle is the only parameter . 

 Denote by $Q^l $ the restriction of $Q$  to the sector of angular momentum $l$

This form can be diagonalized by a Mellin transform $[M_1]$.  

In $ [M_1][M_2] $  the Author proves that there are  constants $ m^{**}_l $ and $ m_l^* $  such that $Q_l$ s positive for $ m>m_l^{**}$ and therefore corresponds to  a self-adjoint operator .

If $ m \leq m^{**}_l$ the form $Q_l$ ceases to be positive and for $ m \leq m^*_l$ it is unbounded bellow. 

Estimates  for these   constants  are $ [C] _1$  $m^*_0 \simeq (13.607) ^{-1}$ and $ m^{**}_0 =  (8,62)^{-1 }$ if the mass of the fermions  is one .

While the two components  represent quadratic forms of  self-adjoint  operators their sum is the quadratic form of an operator that is a priori  symmetric but need not be self-adjoint .

If $ m \leq m^{**} $ the quadratic form is no longer closed. 

The symmetric operator associated to the form can be  "disintegrated" using rotation invariance into a family of self-adjoint operators with negative point spectrum. 

For each extension the space of bound states  is one-dimensional for  $ m^*_ l  < m \leq m^{**} _l $ and infinite dimensional for  $ 0 < m \leq m^*_l $. 

In this latter case the energy of the energies of the bound states accumulate (geometrically)  at minus infinity (Thomas effect).

Recall that the Thomas effect \emph{takes place in} $ { \cal M} $. 

The presence of infinitely many extensions was noticed first by Danilov [Da] (see also [Pa]).

It is  demonstrated  in $ [M_1][M_2]$  by finding in this space infinitely many solutions of an algebraic equation.  

The analysis so far refers to the space $ { \cal M}$.

\emph{To come back to the "physical space" $ L^2 (R^9)$ one must undo the Krein  map}. 

This is not stressed in $[M_1], [M_2]$. 

It is mentioned briefly  that the map back to physical space changes the metric;   due to the difference in metric now the  eigenvalues of the three-body bound states accumulate at zero.  

The use  of weak sequential convergence is not explicitly mentioned.

\bigskip
We now show how to obtain these  thresholds in ${ \cal M}$ by comparing the form we have described with  the form  of the relativistic Coulomb model $ \sqrt{-\Delta} - \frac {C(m) }{ |x| } $ for a suitable function of the mass..

 This  hamiltonian has been thoroughly investigated [D,R], [lY] .
 
 In this model the plurality of extensions for $ m \leq m^{**}_l  $  is clearly seen  as a \emph{ Weyl limit-circle effect}. 

We have noticed that in  $ {\cal M }$ all terms are continuos in $ \lambda $ at zero ; the positivity of $ \lambda$ was essential only in making the embedding in $ {\cal M}$) compact.

Setting $ \lambda = 0$ one has for the potential term if the two identical particles are (spin zero) fermions 

\begin{equation} 
-\frac {1+m } { m } \frac {1} {(p-q)^2 } + \Xi (p,q,\lambda) 
\end{equation}

where $ \Xi $ is a \emph{positive} smooth kernel with  $ \Xi (p,p,0 ) = 0$

Therefore in position space  the  symmetric operator is 

\begin{equation}
2 \pi^2  \frac {m}{m+1} \sqrt {- \Delta} - \frac {1+m}{8 \pi(2m + m^2)} \frac {1}{ |x| } + \tilde  \Xi '
\end {equation} 
 
where $ \Xi' $ is a \emph{positive} operator with locally bounded kernel vanishing on the diagonal.

It follows that the form in $ { \cal M} $ should be compared with the quadratic form of the symmetric operator   

\begin{equation} 
\sqrt {- \Delta} - C(m) \frac {1}{ |x|} 
\end{equation} 

where $ C(m) $ is a suitable positive function of the parameter $m$,

These symmetric operators  have  been studied extensively [lJ][B,R]  as a function of $C(m) $, originally  in the  context of the non relativistic hydrogen atom.

Since they are invariant under rotations, their domain can be decomposed into eigenstates of the angular momentum 

In [B,R] it is proven that for each eigenvalue $l$ of the angular momentum  there are threshold $ M^*_l, \; M^{**} _l $ such that for $ m > M^{**}_l$  the spectrum is absolutely continuous and positive. 

For $ M^* _l< m \leq M^{**} _l$ there is a continuous family of self-adjoint extensions, each with a negative  eigenvalue, and for $ 0 < m \leq M^* _l$ the negative spectrum is pure point and accumulates geometrically  to $ - \infty$ (a Weyl limit circle effect) .

In the latter case the eigenfunctions concentrate at the origin (eigenvalues and eigenfunctions are known explicitly) 

Of course one has  $ M_l^* = m_l^*, \quad  M_l^{**} = m_l ^{**} $. 

It is easy to verify that $ m^{**}_l < 1$  for every $l$.

For equal masses the hamiltonian is positive. 

Recall that these statements \emph{hold true in $ { \cal M} $ }

\section{Two identical bosons and a third particle} 

We now consider briefly the case of two identical bosons of mass 1 and a third particle of mass m.

In the case of two identical boson in $ { \cal M}$ the potential part of the  quadratic form is

\begin{equation}
Q_2(p,q) '= \frac {- \frac{ 2}{1+m} (p.q) } { (p^2 + q^2)^2 - c_f(m)  (p.q)^2 } \qquad c_f(m) = \frac {4}{ (1 + m)^2 }
\end{equation} 
 
 Now  the quadratic form is the sum  of a positive form $ \Xi''$ with  kernel that vanishes on the diagonal and of   $ \sqrt { -\Delta } - \frac {C_b(m) }{|x|}$ where $ C_b (m) > C_f(m)$ ($b$ for bosons).
 
Dropping the positive form $ \Xi'$ one obtains the  "relativistic  atom"  hamiltonian.  

The system is invariant under rotations and can be decomposed in angular momentum states.

Notice that the expectation value of $ \sqrt { - \Delta}$ depends on the angular momentum of the state. 
 
One can again derive the properties of the spectrum from  the spectrum of $H_R$.

Denote by  $M_l^*(b), \; M_l^{**}(b) $ the thresholds in the case of bosons. 
 
 One verifies that  $1 < M_0^*(b)  $ whereas $M_j ^{**}(b)  > 1$ for all $ j \geq 1 $ 
 
 Therefore for two identical bosons of mass 1 which do not interact among themselves and are in contact interaction with a particle of the same mass, in $ { \cal M}$ the Thomas effect is present (there are infinitely many extensions  in the zero angular momentum sector and their energies diverge  geometrically to $ -\infty$). [ B, T].
 
 In physical space this leads to the Efimov effect.
 
\bigskip
\section {N= 3; convergence in physical space}

Recall once more that this analysis is done in $ { \cal M}$, and to draw conclusions relevant for  physics one must come back to physical space inverting the map $ {\cal K}$. 

If the limit quadratic form is positive $ { \cal M}$, it is also positive in physical space (the map preserves positivity).

In this case  the approximating forms in $ { \cal M}$ converge strongly when $ \epsilon \to 0$ and the limit form is strongly closed.

In physical space the convergence is only weak, ad the form is weakly closed.  But for positive forms weak closure implies strong closure therefore if   the limit form is positive, strong resolvent convergence follows. 

If the limit form is not positive, we have seen that there is a continuous family of self-adjoint extensions.

Also here one can consider separately the sectors of fixed angular momentum.

In physical space in each sector, due to the special form of the potential term, the union of the quadratic forms of the self-adjoint extensions   is bounded below and \emph{strictly convex} (the Krein map preserves convexity).

Also the quadratic form of the approximating potentials are bounded below and strictly convex. 

The conditions for Gamma-convergence are satisfied. 

Recall that the Gamma-limit of a sequence of strictly convex forms $F_n$ in a topological function space $ Y$  is the \emph{unique} form $F$  such that for any subsequence the following holds 

\begin{equation}
\forall y\in Y ,\; y_n \to y  \;\;: \;\;  F(y) \leq lim  inf_{n \to \infty} F_n (y_n)  \;\;\; \; \; lim sup_{n \to \infty} F_n (y_n) \geq F(y) 
\end{equation} 

In our case the sequence is parametrized by $ \{\epsilon_n \} $ where is any sequence of positive numbers that converge to zero. 

The limit is independent of the chosen  sequence (we express this independence with the notation $ \epsilon \to 0$).

Therefore  in the singular case for  any choice of the masses  there is in $L^2(R^9) $ \emph{a distinguished extension} (the  Gamma-limit) obtained by Gamma-convergence.

In each angular momentum sector this extension is characterized by having the lowest point spectrum among all possible extensions.

Due to the change in metric the "Thomas spectrum" in $ {\cal M}$  is turned into the "Efimov spectrum" in the physical space $ L^2(R^9) $

And in this space the eigenfunctions are obtained from those in ${ \cal M}$ by convolution with the three-particle Green function. 

\bigskip
We use now the fact that  Gamma-convergence implies strong  resolvent of the associated operators [Dal].

Therefore we have proved 

\bigskip

\emph{ Theorem}

\emph{Contact interactions  between two identical particles and a third  one of the same mass are limits in the strong resolvent  sense of hamiltonians with two body potentials that have support that shrinks to a point. If the identical particles are fermions, the limit hamiltonian is positive.
If the particles  are bosons, the limit hamiltonian is bounded below  and its negative point spectrum is of Efimov type (the sequence of eigenvalues converges geometrically to zero. The eigenfunctions are centered on the triple coincidence point.}

\bigskip
.............

 Resolvent convergence implies, among other things, that bound states converge to bound states. 

Therefore contact interaction are a good tool to find (approximately) the location of the bound states for two body potentials which  are sharply peaked. 

 \bigskip
\emph{Remark 1} 

From the analysis we have made one sees that if the case of bosons, if mass of the third particle converges to zero the three-body point Efimov spectrum diverges to $ - \infty$ in $ {\cal M}$..

\bigskip
\emph{Remark 2}

The "tail" of the Efimov states  (\emph{trimers} in the Physics literature) is  hardly present in a realistic model, which corresponds to $ \epsilon $ very small but not zero .

Therefore what can be seen in experiments are a few  members of the "head" (low energy  states)  of the Efimov sequence; they  should be recognized as Efimov states  for the geometrical scaling property of the energies. 

This states are less affected by the other short range interactions and on them the effect of the zero-energy resonances might  be  more visible.

Three  body Efimov states  have  probably be seen experimentally [Pe], [C,M,P]

\bigskip
.......................

\ \section {Two pairs of identical particles } 

For generic values of the masses and generic symmetry the quadratic form has a complicated structure. 

We consider here only the case of  two pairs of particles of equal mass, either fermions or bosons; in this case  the quadratic forms are defined on a space of functions of two variables.

Again we start  the analysis in ${ \cal M}$.

The kinetic term  is again  $ \sqrt H_0$ , the potential term  is the convolution of the four particle Green function with the distributional potential. 

The denominator in the convolution  can be decomposed in different ways in the sum of squares corresponding to the cases in which a particle   does or does not belong to a same pair. 

In spatial coordinates this allows to isolate first order polar singularity in different positions.

Depending on the statistic (bosons or fermions) the different components can be either all negative (for the bosonic case; recall that the potential is negative) or have alternate signs.

As a consequence the total quadratic has a kernel that  may be positive (fermionic case) or it may have a negative unbounded component with  Coulomb singularities  located in different positions.
 
 \bigskip
 Consider  first the  system of two pairs of identical scalar fermions of mass one in contact interaction. 
 
 One can equivalently consider a system of identical spin $ \frac {1}{2} $ fermions.

The generalization to  $N$ identical spin $ \frac {1}{2} $ fermions will describe the unitary gas. 

In $ { \cal M}$  the quadratic form  is  the sum a term $C_0$ which represents the kinetic part of form  minus three forms $C_1, \; C_2 \; C_3 $.

The explicit expressions of these forms in momentum space were known to R.Minlos (private communication).

 $C_1 $ and $C_2 $ are the images in $ {\cal M}$ of the convolution of the four-particle Green function with  two delta singularities of the potential between two identical particles and a third one

\begin{equation}
(\phi, C_1\phi)  = (\phi, C_2 \phi) = \int dk ds d w \bar \phi (k,w) \frac {\phi (s,w) + \phi (k,s) }{ k^2 + s^2 + w^2 + (k,s) + (k,w) + (s,w) }  
\end{equation}

As in the tree particles case, when written in space coordinates they have a negative Coulomb type singularity in different variables related to the possible triples.

Since they refer to pairs of identical spin $ \frac {1}{2}$ fermions the contribution are identical. 
 
Contributions  $C_1,\;C_2$ refer to the three particle case, i.e. two identical fermions  and another particle with the same mass, that we have already considered.

They differ from the expressions found in the case of three particles because for the Krein map we have used here the  resolvent of the free four particle system instead of the one for the three particle case. 

This difference disappears upon coming back to physical space. 
 
 \emph{In physical space in this contribution  the presence of a fourth particle is irrelevant} . 
 
 We stress this fact because it means that in the formation of  \emph{trimers} the other body (or bodies) \emph{plays no role}.

\emph{The same will be true in the case of $N$ particles, $ N > 4 $.}

\bigskip
$C_3$  is a genuine four particle term which is not present in the three-particle sector. 

\emph{It represents an effective interaction between the barycenters of the two pairs}   

The corresponding  quadratic form in { \cal M} is

\begin{equation}
(\phi,C_3 \phi)  = - \int dw ds dk \frac {\bar \phi (k,s) \phi (w- \frac {k+s}{2} , -w-\frac{k+s}{2})} { w^2 + \frac {3}{4} (k^2 +s^2)   + \frac {1}{2} (k,s) } 
\end {equation} 

 It has a simpler when written as a function of the difference of coordinates of the barycenters of the two  pairs. 
 
 In these  coordinates it is the image  \emph{in the four particle sector of } ${ \cal M}$ of an "effective" contact interaction between the barycenters of the two pairs. 
 
\bigskip
Notice that only \emph {pairs of particles}  enter in this term.

The problem is reduced to the three particle case..

The form can be decomposed into a symmetric and antisymmetric part under interchange \emph{of the two pairs}. 

Only the kinetic energy contributes to the antisymmetric part; this part is positive. 

The symmetric part can be decomposed  as  sum of a term which is symmetric and a term which is antisymmetric under interchange of the elements \emph{in one of the pairs}.

For fermions the symmetric term is positive (only the kinetic energy contributes) .

From the analysis of the three particle problem it follows that also the antisymmetric term of is positive..

Therefore in the  case of two pairs of fermions the quadratic form is positive.

Since the Krein   map is positivity preserving the same is true in physical space.

 We  have  proved

\bigskip 
\emph{Theorem }

\emph{ The operator associated to a system two pairs of identical fermions  in contact interaction is a  positive self-adjoint operator in $L^2(R^{12})$. 
Its hamiltonian is  the limit, in the strong resolvent sense, of a sequence of approximating hamiltonians with potentials of decreasing support.}

 \bigskip
 ..............
 
In [M.P] positivity was conjectured with the aid of a computer.

\bigskip
Consider now the case of two pairs of identical bosons.

Also here there are four terms $ C'_k \; k=0,1,2,3 $. 

As in the fermionic case, $C'_1$ and $C'_2$ \emph{and describe a three-particle system}.

$C'_3$ \emph{is negative}.  

It is different from zero only in the four -particle sector and has a first order pole in the barycenter of the two pairs. 
 
It can  be decomposed in a symmetric and antisymmetric part under interchange of the two pairs. 

The antisymmetric part is positive; it represents the contact  interaction of two fermions with a boson.

The symmetric part represents the (contact) interaction between two bosons (with reduced mass) with the barycenter. 

From the results in the case of three bosons one derives that this term is not positive, convex and has a $ -\frac {1}{ |x_i - x_b| }$ singularity.

It can be decomposed using rotation invariance. Each component is strictly convex. 

It leads  to a one-parameter family of self-adjpint extensions.  

When lifted to physical space,   under Gamma-convergence it represents a unique self-adjoint operator; its point spectrum  represents an Efimov  sequence of \emph{quadrimers}.

\bigskip
\emph{Theorem} 

\emph{ A system of  two pairs of identical bosons in two body contact interaction has an infinite number of three body bound states (trimers) and of four-body bound  states  (quadrimers); both show the Efimov effect. The hamiltonian of the system is the limit in strong resolvent sense as $ \epsilon \to 0$ of hamiltonians  with potentials of radius   $ \epsilon$ and $L^1$ norm independent of $ \epsilon$.}

\bigskip
.......................................

As for the three-body case, one expects to see experimentally only the \emph{head} of the Efimov tail. 

Four-body Efimov states have been reported [C,M,P][Ba,P].

\section{N-body sistems}

We have considered so far the cases $N=3 $ and $ N=4$.

Consider now the case of $N$ identical  spin $ \frac {1}{2} $ fermions or $N$ identical bosons  in contact interaction.

We have noticed that the negative part of the spectrum in $ {\cal M} $ is \emph{entirely due to the conspiracy between pairs of contact interactions}.

This effect is \emph{independent from the presence of other particles}. 

From an analytic point of view, this is a consequence of the fact that  in $ { \cal M}$ the "kinetic part" is a first order differential operator and its Coulomb capacity is zero.

By additivity, the hamiltonian of a system of $N$ identical spin $ \frac {1}{2} $ fermions is positive.

Therefore 

\bigskip
\emph{Theorem}

\emph{Under contact interaction  a gas of $N $ spin $ \frac {1}{2} $ fermions   is stable i.e. its hamiltonian is a positive self-adjoint operator.
It is the limit, in strong resolvent sense, of interactions through potentials $V^\epsilon _{i,j} (|x_i-x_j|)$ that scale as $V^\epsilon _{i,j} (|y|) = \frac {1} {\epsilon^3} V_{i,j} ( \frac { |y| }{ \epsilon}) , V_{i,j}  \in L^1 (R^3)$}

\bigskip
...........................

For identical bosons in the space $ {\cal M}$ the potential corresponding to a contact interaction of each pair   has a   $ \frac {1}{ x_i -x_k}$ singularity. 

In $ \cal M$ there are infinitely many extensions and for each of them  there is an infinity of bound states for which  the Thomas effect is present.

But also in this case \emph{the conspiracy  between two contact interactions  is independent from the presence of other particles}.

Therefore in physical space the hamiltonian is bounded below by $ - c N$ and  shows the Efimov effect. 

The positive constant c depends on the masses and on the strength of the boundary potential.
 
\bigskip
\emph{Theorem}

\emph{The hamiltonian of a system of $N$ identical bosons in contact interaction is bounded below by $ - c N$  where the constant $c$ depends on the mass and on the strength of the interaction (the coefficient of the delta function at the boundary). 
The negative part of the spectrum is composed entirely of Efimov trimers and quadrimers . 
The hamiltonian is limit, in strong resolvent sense, of hamiltonians with (negative) two-body potentials $V^\epsilon_{i,j} (|x_i- x_j|)  = \frac {1}{\epsilon^3}V_{i,j} ( \frac {|x_i-x_j| }{ \epsilon} $}).

\bigskip
..........................

\bigskip
\emph{Remark}

One can also consider a system composed of $N$ fermions and $M$ bosons, all with the the same mass, interacting through contact interactions.

Only interactions among tree or four of the particles are relevant. 

It is then easy to see that for $M=1$ the system is stable independently of $N\geq 2$; if $M=2$ and $ N \geq 2 $  the  system has an infinite number of bound states and the Efimov effect is present. .

For an outlook on experimental and theoretical results on the $N$body  problem one can consult [C,M,P] [C,T] [Pe].

\bigskip
 .............................

\section{ Krein  resolvent formula for contact interactons}

In quantum mechanics the Krein  formula for the difference of two resolvents   is used to transform a relation between unbounded operators into a relation between two families of bounded operators; it is mostly used to study convergence. 

For regular two-body potentials if we set $ H_\epsilon = H_0 + V_\epsilon  $   the Krein  map ${\cal K}_\lambda $ is defined by duality  on quadratic forms by  

\begin {equation} 
 {\cal K}_\lambda^\epsilon  (W ) =( \frac { 1}{ H_0 + \lambda})^ { \frac {1}{2} } V_\epsilon   ( \frac { 1}{ H_0 + \lambda}) ^ { \frac {1}{2} } 
\end{equation}

where $ V_\epsilon $ is the quadratic form of the potential term. 

For smooth potentials the Krein formula for the difference of two resolvent families is

\begin{equation}
\frac {1}{ H_\epsilon  +\lambda } - \frac {1}{H_0 +\lambda } = \frac {1}{H_0 +\lambda }  W_{\lambda ,\epsilon}      \frac {1}{H_0+\lambda } 
\end{equation} 

where   $  W_{\lambda ,\epsilon} $ is the Krein  kernel and $H_\epsilon $ is the total hamiltonian .

For contact interactions elements $ \psi $ in  the domain of $H_\lambda $ can be written [A,S] as $ \psi = \phi + \xi$ where $ \phi $ is an element the form domain of $H_0$  and $ \xi $ square integrable but more singular (it belongs to the space ${ \cal M}$).

Under this decomposition one has

\begin{equation}
(H_\epsilon + \lambda) \psi = (H^0+ \lambda) \phi _\lambda 
\end{equation} 

and the quadratic form of $H$ can be written [A,S] for $ \lambda =0$, as  

\begin{equation}
 \Phi (\psi_1, \psi_2) = \Phi_0 (\phi_1, \phi_2)  + \Xi (\xi_1,\xi_2) 
 \end{equation} 
 
 where $ \Phi_0$ is the quadratic form of $H_0$ and  $ \Xi $ is a bilinear form  in $ {\cal M}$.

For $ N \geq 3 $ the Krein kernel for the approximating hamiltonian converges in the limit   $ \epsilon \to 0$  .  

We can therefore  study  the resolvent  before the limit $ \epsilon \to 0$ by lifting it   to ${ \cal M }$  where the "boundary potentials"   have a  finite $L^1$ norm . 

In this space it is easy to verify  convergence. 

Also the Krein kernel converges. 

In the regular case the convergence holds also in the physical space because the quadratic form is positive. 

Also in the singular case by Gamma-convergence there is a unique limit operator. 

Resolvent convergence occurs away from the spectrum  of this operator. 

 Using $ { \cal K_\lambda } $ for $\lambda > 0$ the Krein  formula can be written  

\begin{equation} 
\frac {1}{H + \lambda }  -  \frac {1}{ H_F + \lambda  }  = {\cal K}_\lambda (  {\cal K}_\lambda ( W_\lambda ) ) 
\end{equation} 

where $ W_\lambda = \lim_{\epsilon \to 0 }W_{\lambda, \epsilon}  $ exists  due to the smoothing properties of $ {\cal K_\lambda } $.

For a general study of the Krein resolvent formula for self-adjoint extensions see [P,R].

\bigskip
\emph{Remark}

For contact interactions there is a natural relation between the Minlos  kernel and the quadratic form that characterizes the extension.

To see this, notice that the domain is the sum of a part that belongs to the domain of the free laplacian and a more singular part (it correspond in electrostatics  to the contribution of the charges). 

The relation is shown by writing in two different ways the energy form.

Let $H$ be the self-adjoint extension that represent the contact interaction.

 Choose $ \lambda $ in such a way that  $ H_\lambda = H +  \lambda I $ is invertible

Introducing the Krein kernel $ W_\lambda $ defined by

\begin{equation} \frac {1}{  H + \lambda} = \frac {1}{ H^0 + \lambda} + \frac {1}{ H^0 + \lambda} 
W_\lambda \frac {1}{ H^0 + \lambda}
\end{equation} 

one has  

\begin{equation}
(H_\lambda \psi, \frac {1}{H_\lambda } H_\lambda  \psi) =  ( H^0_\lambda \phi , \frac  {1}{ H_\lambda } H^0_\lambda \phi) = (\phi, H^0_\lambda \phi) + (K_\lambda  \phi, K_\lambda    \phi) 
\end{equation}

This relation should be compared with (17) [A,S].

\section{  Further comments }

We conclude  with an outline of possible developments. 

Notice the similarity of  the  scaling  of the potential with parameter $ \epsilon$ (the inverse of the scattering length) with  the scaling $ V_N(x) = N^{-1}(N^3 V(\frac {|x|}{N}))$ which is used in the study of the fluctuations of N-particle quantum systems, for $N$ very large, in the  Gross-Pitaewskii regime (see e.g. [B,C,S]) .

The coefficient $N^{-1}$ is introduced in [B,C,S] to balance the kinetic and the potential terms. 

In our case the extra factor $ N^{-1} $ is introduced to take care of  the lower bound in the spectrum of a bosonic $N$-body system with contact interactions. 

This permits to see the landscape of the bound states.

If the gas of particles is  very diluted, "most of the non-free configurations" correspond to interaction of  two particles. 

A priori the parameters $ \epsilon $ (inverse of the scattering length of the potential) and $\frac {1}{N}$ are \emph{ independent parameters}.

For a system with al large number $N$ of particles it is natural to take $\epsilon   \simeq  \frac {1}{N}$. 
This allows to use both Fock space techniques [B,C,S] and techniques related to the Schr\"odinger equation for a many body system with regular potential of very small (contact  interactions).

The  choice of consider first the limit $ \epsilon \to 0$ and then the limit of very large $N$, has some advantages. 

It permits to prove that the only bound states  present for any value of $N$ are three- and four-body bound states.Their total energy is bounded uniformly by $ - cN$.

Recall that the explanation  given in [B,C,S] is that this factor is needed to give equal  weight to the kinetic and the potential term in the limit $ N \to \infty $.

\bigskip
Consider now  a system of N  identical bosons  of mass one  in contact interaction and restrict attention to the case in which two particles can be very close (in the limit, in contact) but  the system is so diluted   that the probability that find another particle nearby is negligible. 

\emph{The equation for contact  interaction is defined also for a system of two isolated particles}. 

If we assume that   the probability of interaction of the two particles  is proportional to the probability that they be simultaneously present, contact interaction leads in physical space to the non linear equations for the wave functions of the two particles 

$$
i \frac {\partial}{\partial t} \phi_1  (t,x)= \frac {1}{m_1} \Delta \phi_1 (t,x) + c |\phi_2  (t,x) | ^2 \phi_1(t,x) \qquad 
$$
\begin{equation}
i \frac {\partial}{\partial t} \phi_2  (t,x)= \frac {1}{m_2}  \Delta \phi_2 (t,x) + c |\phi_1  (t,x) | ^2 \phi _2(t,x) 
\end{equation}

where $c$ is a  coupling constant.

These are the equations of two  particles in contact interaction. 

 Since contact  interactions are strong resolvent limit of interactions through  short range potentials when the range goes to zero,  \emph{ the solutions of these effective equations} , if unique, are strong limits, as $\epsilon \to 0$  of  the solutions of the equation for the system of the two selected particles under the assumption  on the initial data that the probability distributions are identical.  

Setting  $ |\phi_1 (x) |^2 = |\phi_2 (x) |^2 $ one can regard these equations as effective equations for  a gas of  identical  particles  if the gas is very  diluted and the interaction is of  range so small that one can omit the interactions of more than two particles.

The resulting equation \emph{for each particle in the pair} is

\begin{equation}
i \frac {\partial}{\partial t} \phi  (t,x)=- \frac {1}{m}  \Delta \phi (t,x) + c |\phi  (t,x) | ^2 \phi (t,x) 
\end{equation}

The equation is meant to describe the dynamics of the one-particle marginals for a gas  of identical particles  if the gas is very  diluted and the  range of the interaction is so small that it can be described as a contact  interaction. 
  
On the other hand, the solutions to this equation may be regarded  as describing, in the contact interaction limit,   the solutions of a  \emph{linear equation} for a pair of identically distributed  particles. 

Notice that the resulting equation is linear  \emph{for each member of the pair}, but \emph{it is the pair structure that is relevant}.

\section {Conclusions}

We have proved that for a Schr\"odinger system of three  or more particles in contact interactions  the hamiltonians are self-adjoint operators which are constructed considering a natural extensions of $\hat H_0$ , a symmetric operator defined on functions with support away from a subset of coincidence planes  $ \Gamma_{i,j} $ (defined by $ x_i = x_j$).  

These operators are limits, in the strong resolvent sense, of Schr\"odinger hamiltonians with two-body potentials with smaller and smaller support and  constant  $L^1 $ norm. 

If there are bound states this is proved by Gamma-convergence [Dal]. 

The negative (point-) spectrum  of the Hamiltonian for contact interactions is completely determined by the structure of the three-body and four-body subsystems. 

Functions  in the  continuous spectrum of the contact hamiltonian satisfy  T-S boundary conditions at $ \Gamma_{i,j}$. 

The bound states are due to conspiracy of two-body contact interactions;  three and four-body bound states are Efimov states.

\bigskip
Acknowledgments 

Comments and constructive criticism by several friends at an early stage of the research were very helpful and are gratefully acknowledged. 

I'm grateful to R.Minlos for valuable correspondence and  to K.Yajima for the very warm hospitality at Gakushuin University

\bigskip
\centerline{ \emph{References} }

\vskip 4 pt \noindent
[A] S.Albeverio, F.Gesztesy, R.Hoegh-Krohn, H-Holden  Solvable models in Q. M. AMS 2004
\vskip 4 pt \noindent
[A, K] S.Albeverio, P. Kurasov   Lett. in Math. Physics  41 (1997)  79-92
\vskip 4 pt \noindent
[A,S] A.Alonso, B.Simon J. Operator Theory 4 (1980) 251-270
\vskip 4 pt \noindent
[B] M.Birman Math. Sb. N.S. 38 (1956) 431-480 
\vskip 4 pt \noindent
[B,P] H. Bethe A.Peierls Proc. Royal Soc. 148 (1935) 146-164
\vskip 4 pt \noindent
[B,M,P]  J.Brasche, M.Malamud, H.Neidhardt  Int.Eq. Op. Theory 43 ("002) 264-289
\vskip 4 pt \noindent
[B,C,S] C.Boccato, S.Cenatempo, B Schlein Ann. Henry Poincare' 18 (2017) 113-191
\vskip 4 pt \noindent
[Ba,P]  B.Bazak, D.Petrov Phys. Rev Lett. 119  (2017) 083002-083006
\vskip 4 pt \noindent
[B,T] G.Basti, A.Teta  arXiv 1601.08129 (1916) 
\vskip 4 pt \noindent
$[C_1]$ M.Correggi et al  Rev.Math.Phys. 24 (2012) 1250017-32 
\vskip 4 pt \noindent
$[C_2]$ M.Correggiet al. Mathematical Physics, Analysis, Geometry 18(2016) 1--36
\vskip 4 pt \noindent
[C,M,P] Y. Castin, C.Mora, L.Pricoupenko PRL 105 (2016) 2232011-4
\vskip 4 pt \noindent
[C,T]  Y. Castin, E. Tignone Physical Rev A 84 (2011)  062704- 062720
\vskip 4 pt \noindent
[D,M,S,Y] G.F.Dell'Antonio, A.Michelangeli, E.Scandone, K.Jajima Ann Inst.H  Poincare' 19(2018) 283-322
\vskip 4 pt \noindent
[Da]  G.S.Danilov Sov. Phys. JETP 13 (1961) 648-660) 
\vskip 4 pt \noindent
[Dal]  G.Dal Maso Introduction to $ \Gamma$-convergence Progr Non Lin. Diff Eq. 8, Birkhauser (1993)
\vskip 4 pt \noindent
[D,R] S.Derezinski, S.Richard ArXiv:1604.03340 
\vskip 4 pt \noindent
[D,H,M]  V.Derkach, S.Hassi, M.Malamud   arXiv:1706.07948 
\vskip 4 pt \noindent
[D,M]  V. Derkach, M.Malamud J. Funct. Analysis 95 (1991) 1-95
\vskip 4 pt \noindent
[D,R] J.Dimock , S Rajev arXiv math-ph 0403006
\vskip 4 pt \noindent
[E] V.Efimov Nucler Physics  A 210 (1971) 157-186
\vskip 4 pt \noindent
[E,T] F.Erman, O.Turgut, J.Phys. A (2010) 335204
\vskip 4 pt \noindent
[K] M.G.Krein Rec Math. (Math. Sbornik)   N.S. 20 (1962)1947 431-495
\vskip 4 pt \noindent
[K,R] K.Kowalsy , J.Rembielinski Phys. Rev. A 84 (2011) 012108
\vskip 4 pt \noindent
[K,S] M.Klaus, B.Simon  Ann. Inst. Henry Poincare  XXX, (1979) 83-87
\vskip 4 pt \noindent
[lY] A. Le Yaouanc, L. Oliver, J-C Raynar J. of Mathematical Physics 38 (1997) 3998-4012
\vskip 4 pt \noindent
[M,F]  R-Minlos, L.Faddeev   Sov. Phys. Doklady 6 (1972) 1072-1074
\vskip 4 pt \noindent
[M,P] A.Michelangeli, P.Pfeiffer   J. PhysicsA  49 (2016) 105301-105331 
\vskip 4 pt \noindent
$ [M_1] $ R.Minlos    Moscow Math. Journal  1 (2011) 111-127
\vskip 4 pt \noindent
$ [M_2]$ R.Minlos  Moscow  Mat. Journal 14 (2014) 617-637
\vskip 4 pt \noindent
[P] A.Posilicano Journ. Functional analysis  183 (2011) 109-147
\vskip 4 pt \noindent
[P,R]  A.Posilicano , L.Rinaldi J.Phys. A Math.Theor. 42 (2009) 015204 
\vskip 4 pt \noindent
[ Pa] B.S.Pavlov  Math. Sbornik 64 (1989) No 1 
\vskip 4 pt \noindent
[Pe]  D.S. Petrov  Physcal Rev Letters 93 (2004) 143201- 143204 
\vskip 4 pt \noindent
[S,B] E.Salpeter, H.Bethe Phys Rev 84 (1952) 1232
\vskip 4 pt \noindent
 [S,T]  G.V.Skorniakov, K. A. Ter-Martirosian Soviet Physics JETP 4 (1957) 648-661

\bigskip
\section{Appendix : contact interaction vs point interaction}

If the two-body potential  has a zero energy resonance \emph{a different }  hamiltonian of "zero range interaction"  was introduced in [A] ; the resulting system is  called  \emph{point interaction}.

The hamiltonian  is obtained as limit of of $ H_0 + V_\epsilon$ where  
$ V_\epsilon (|x|)=  \epsilon^{-2}  V(\frac {|x|} {\epsilon}) $ and the potential $V(|y|)$ has a zero energy resonance. 

Notice that  in the limit the $L^1$ norm vanishes.
 
 It is proved in [A] that this  Hamiltonian is self-adjoint. 
 
 The proof requires an accurate balancing in the Krein resolvent formula of the divergence due to the resonance and the vanishing of the $L^1$ norm of the potential.

\bigskip
In order to see the relation between contact interaction and point interaction, it is convenient to remark  that  zero energy resonances represent  directions in which the wave function has a $\frac {1}{|x_i-x_j|}$ behavior. 

This is the same behavior as that at $ \Gamma$ for contact interactions. 

Therefore in  conformal coordinates (the laplacian in dimension 3 is conformal covariant), resonances can be  seen as  contact interaction \emph{at infinity  between a pair of particles}.

In this sense the Efimov effect for contact interactions  has the same origin (conspiracy of contact interactions) as  the effect with the same name in low energy nuclear physics [A.S] (conspiracy of zero energy resonances). 

In the case of two particles  of equal mass which are in contact interaction and have a zero energy (Feshbach)  resonance,  weakly bound pairs (Cooper pairs) are the counterpart of quadrimers. 

\bigskip
A resonance  between \emph{two particles} is due to a two body potential with  infinite  scattering length. 

Denoting by $ \epsilon $ the inverse of the scattering length, the resonance between two particles provides a factor  $ \frac {1}{\epsilon^3} $ that makes the volume integral finite. 
 
If one of the particle has very large mass $ \frac {1}{\epsilon  }$ the kinetic energy becomes small for a large spectrum of momenta and it is convenient to scale the masses by a factor $ \epsilon^{-1}$.

This does not affect the resonances and changes the scaling of the potential; it now scales as $ V^\epsilon (|x|)  = \frac {1}{ \epsilon^2}V( \frac {|x|}{ \epsilon})$. 

On the new scale he  system consists of a particle of very small mass in contact interaction with a particle of mass one in presence of a zero energy resonance (zero energy resonances are scale invariant), 

In the limit $ \epsilon \to 0$ the particle of mass $ \frac {1}{\epsilon}$  can be considered fixed at the origin and one recovers in the limit the hamiltonian of point interaction described in [A]. 

In physical space the hamiltonian obtained by  Gamma-convergence is bounded below and has Efimov spectrum; but notice that before the limit the eigenvalues are spaced of order $ \epsilon$ since the potential is reduced by a factor $\epsilon$.

When the spectrum has a negative part it is bounded  below and each point is an accumulation point. 

\emph{Therefore on the negative axis the spectral measure  is continuos but not absolutely continuous} and has a singularity at the the origin (the onset of the continuous spectrum).

If the hamiltonian is positive the spectral measure has a singularity at zero(this is due to the resonance). 

This "explains" the the "anomalous" mapping properties of the Wave operator for point interactions [A].

\bigskip
A simpler way to find a relation  between \emph{contact}  and \emph{point interactions} is to notice that  zero energy resonances have   \emph{ long time scale effects}.

Consider a system composed of particle $A$ of mass one and particle $B$ of mass $ \frac {1}{ \epsilon}$ and particle interacting   through a potential $V_\epsilon   = \frac {1}{ \epsilon^3}V(\frac {|x|}{\epsilon})$ , $ V(x) \in L^1 (R^3) $ that admits a zero energy resonance (this is true independently of $ \epsilon$).. 

The dynamics is uniquely defined for arbitrary  small values of $ \epsilon$.

We scale time setting $ \tau = \epsilon  t $ so that on the $ \tau $ time scale the displacement  of particle $A$ is  of order $ \epsilon^{-1 } $.

In the limit $ \epsilon \to 0$ particle $B$ can be regarded as fixed at the origin.

With  this approximations \emph{and on the new time scale}  the Schr\"odinger equation  reads 

\begin{equation} 
 i \frac {\partial }{ \partial \tau } \phi = - \Delta \phi + \epsilon ^{ -2 } V (\frac {| x |}{\epsilon})  \phi 
\end{equation} 
 
In the absence of zero energy resonances the dynamics converges in the limit  $ \epsilon \to 0$ to free motion. 

If a zero energy resonance is present,  in the limit $ \epsilon \to 0$ \emph{and in the new time scale}  the dynamics is given  the Schr\"odinger equation  for a particle of unit mass interacting through a   \emph{point interaction} with a particle fixed at the origin. 

Therefore  point interaction with a particle fixed at the origin can be regarded as an approximate description  of  the asymptotic (in time) dynamics of a particle of very small mass interacting  with a particle of large mass through a potential of very short range that gives rise to a zero energy resonance. 

Notice that after a time of order $ \epsilon^{-1} $ the motion is  almost free (ballistic) if the two-body potential havs  no zero energy resonance and is described by a point interactions if there is a zero energy resonance.  

If one recalls that the spectrum of point interaction is singular at  the origin  this justifies  the mapping properties of the Wave Operator for point interaction [DMSY]. 

\bigskip
\emph{Remark}

We add here a comment on the relation between our procedure and an analysis described in [E,T] using
 "heat kernel regularization".

This is a procedure that regularizes by introducing in the expectation value of the operators a gaussian regularizing factor $ e^{ - \tau (H_0 + \lambda)} $ and  taking the limit $ \tau \to \infty$.

In a quadratic form analysis this  map is precisely what we have called Krein   map. 

One has now to come back to the "physical space" . This must be the meaning of the "renormalization procedure" in [E,T]

\end{document}